\documentclass[aps,prl,twocolumn,amsmath,amssymb,10pt]{revtex4-2}

\usepackage{graphicx}
\usepackage{bm}
\usepackage[usenames,dvipsnames]{color}
\definecolor{altncolor}{rgb}{0,0,0.8}
\usepackage[colorlinks=true, linkcolor=green, anchorcolor=altncolor,
citecolor=altncolor, filecolor=altncolor, menucolor=altncolor,
urlcolor=altncolor]{hyperref}

\begin{document}

\title{Doppler sensitivity and optimization of Rydberg atom-based antennas}

\author{Peter B. Weichman}

\affiliation{FAST Labs$^{TM}$, BAE Systems, 600 District Avenue, Burlington, MA 01803}

\date{\today}

\begin{abstract}

Radio frequency antennas based on Rydberg atoms can in principle reach sensitivities beyond those of any conventional wire antenna, especially at lower frequencies where very long wires are needed to accommodate the growing wavelength. This paper presents a detailed theoretical investigation of Rydberg antenna sensitivity, elucidating parameter regimes that could cumulatively lead to 2--3 orders of magnitude sensitivity increase. Of special interest are three-laser setups proposed to compensate for atom motion-induced Doppler spreading. Such setups are in indeed shown to be advantageous, but only because they restore sensitivity to the \emph{expected} Doppler-limited value, removing significant additional off-resonance reductions.


\end{abstract}

\maketitle

\begin{figure}

\includegraphics[width=2.75in,viewport = 0 0 410 390,clip]{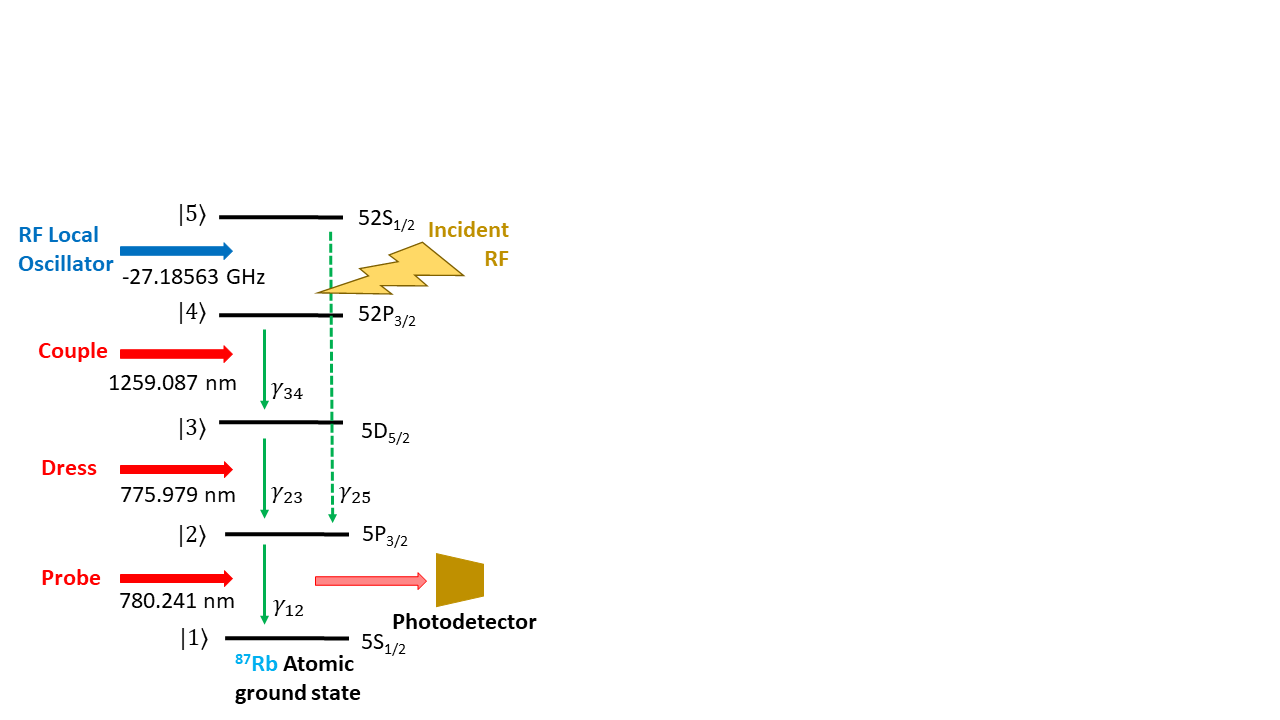}

\caption{Example three-laser, five-state $^{87}$Rb Rydberg atom setup. In combination, Probe, Dress, and Coupling beams promote the atom from its ground state $|1\rangle$ to a selected $n=52$ Rydberg state $|4\rangle$. An additional vapor cell RF local oscillator \cite{NIST2019a,NIST2019b} couples the latter to a second Rydberg state $|5\rangle$ (lower in energy in this case). The incident field, with small frequency difference $|\omega_\mathrm{in} - \omega_\mathrm{LO}| \ll \omega_\mathrm{LO}$, is superposed upon the latter, and through resonant tuning influences the Probe beam transmission $P_\mathrm{EIT}$. The dominant single photon decay paths are also shown. The values for this example are $(\gamma_{12}, \gamma_{23}, \gamma_{34}, \gamma_{25})/2\pi = (6606.50, 646.18, 1.64, 1.09)$ kHz.}

\label{fig:rysetup}
\end{figure}

Rydberg atoms are formed by exciting a ground state electron into a very high orbit. The resulting hydrogen-like atomic state can have (depending on the exact excited state) an electric dipole moment scaling with principal quantum number $n$ as $p_\mathrm{Ry} \sim e a_0 n^2$ where $a_0$
is the Bohr radius. For $n = O(10^2)$ this implies a potential $O(10^4)$ increase in the coupling strength of the atom with an incident external electric field ${\bf E}_\mathrm{in}(t)$.

Designing practical devices requires a number of advances that have been accumulating over the past 30+ years \cite{HFI1990,FIM2005,SK2018,Michigan2019,MITRE2021}. Detection sensitivity actually requires two nearby Rydberg states with energy difference close to the resonance condition $\Delta E_\mathrm{Ry} = \hbar \omega_\mathrm{in}$ where $\omega_\mathrm{in}$ is the incident field center frequency, constrained also by the photon unit angular momentum. The ``frequency tuner'' part of the setup hence requires choosing different state pairs to obtain near-resonance. There are many such pairs available, enabling unprecedented coverage from near-DC to to THz regimes \cite{Durham2017,JC2020,ARL2020}.

The Rydberg state population is maintained by a finely tuned laser setup, exciting the atom via a selected sequence of transitions. The focus here is the 3-laser setup, Fig.\ \ref{fig:rysetup}, enabling Doppler control \cite{foot:3laser}, significantly extending a recent 2-laser setup analysis \cite{ARL2021}. The combination of laser, local oscillator (LO) \cite{NIST2019a,NIST2019b}, and incident field illumination creates a coherent superposition of core and Rydberg states. When resonantly tuned, this state is highly sensitive to ${\bf E}_\mathrm{in}$, permitting its readout through variation of the Probe beam transmission $P_\mathrm{EIT}(t) = P_\mathrm{ph}(t)/P_0$ where $P_0$ is the laser power and $P_\mathrm{ph}$ the power at the photodetector \cite{HFI1990,FIM2005,SK2018,Michigan2019,MITRE2021}. The effect is known as electromagnetically induced transparency (EIT). The Rydberg coupling is quantified by the Rabi frequency
\begin{equation}
\Omega_\mathrm{Ry}(t)
= \frac{1}{\hbar} {\bf E}_\mathrm{Ry}(t) \cdot {\bf d}_\mathrm{Ry}
= \frac{1}{2}[\Omega_\mathrm{LO} + \Omega_\mathrm{in}(t)]
\label{1}
\end{equation}
with (large) transition dipole moment ${\bf d}_\mathrm{Ry}$ \cite{FIM2005} and ${\bf E}_\mathrm{Ry} = {\bf E}_\mathrm{LO} + {\bf E}_\mathrm{in}$ the total field in the narrow band influencing the Rydberg transition. The LO multiplier $e^{-i\omega_\mathrm{LO} t}$ has been factored out, and $\Omega_\mathrm{in}(t)$ may then be thought of as a narrow band communication signal centered on the (small) frequency difference $\Delta \omega_\mathrm{in} = \omega_\mathrm{in} - \omega_\mathrm{LO}$ \cite{NIST2019c,NIST2022,BBNTS2022}.

\begin{figure}

\includegraphics[width=3.0in,viewport = 0 0 960 450,clip]{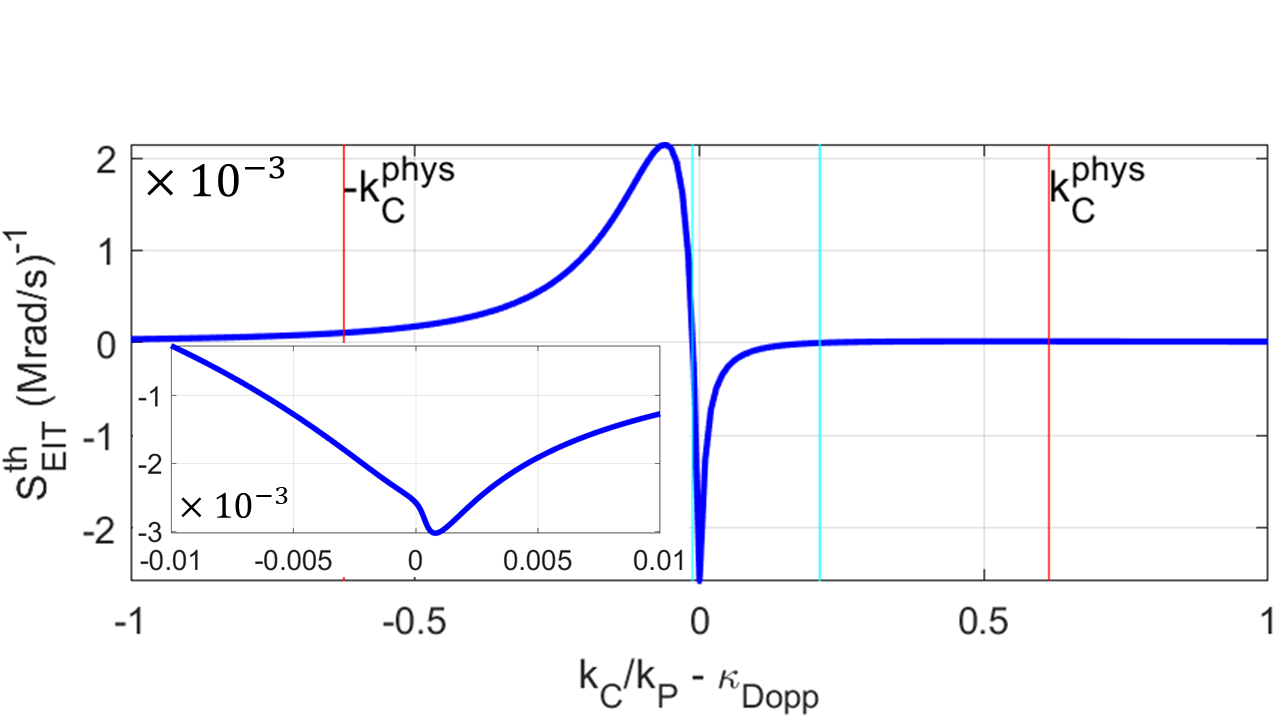}
\includegraphics[width=3.2in,viewport = 0 0 940 450,clip]{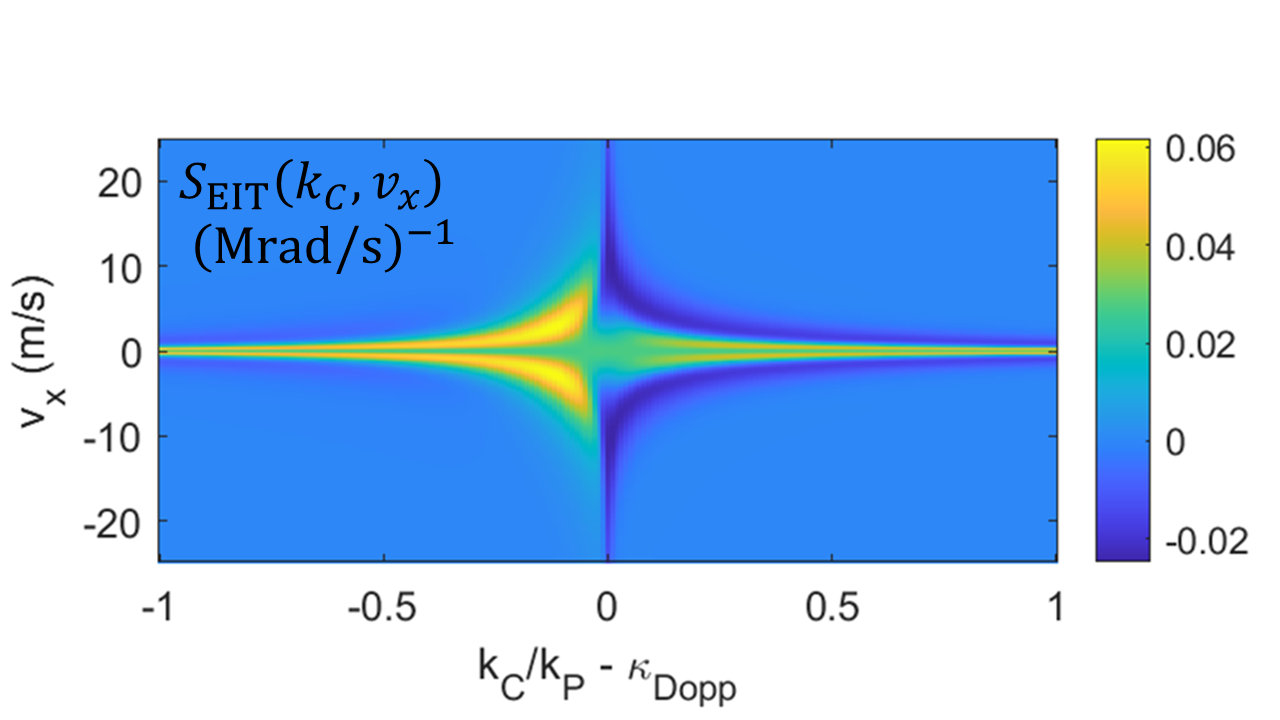}

\caption{\textbf{Top:} Sensitivity $S_\mathrm{EIT}^\mathrm{th}(T=300\ \mathrm{K})$ vs.\ Doppler detuning for the 1D setup: ${\bf k}_D$ antialigned with ${\bf k}_P$, ${\bf k}_C$. Vertical cyan lines mark zero crossings. Here $k_C$ is varied through the Coupling parameter $\Delta_C + k_C v_x$ at fixed $\Delta_C$. Using the parameters in Fig.\ \ref{fig:rysetup}, the true physical sensitivity ($k^\mathrm{phys}_C/k_P = 0.6197$; vertical red lines) is $< 1$\% of its value at the Doppler point $\kappa_\mathrm{Dopp} \equiv k_D/k_P - 1 = 0.00549$. Experimentally motivated parameters are $L = 1$ cm, $(\Omega_P,\Omega_D,\Omega_C,\Omega_\mathrm{LO})/2\pi = (2.7, 3.9, 3.1, 0.85)$ MHz and all $\Delta_\alpha = 0$. \textbf{Inset:} Expanded view of the very narrow negative-going main peak. The $O(10^{-3})$ scale is consistent with the underlying $O(1\ \mathrm{MHz})$ spectral linewidths. \textbf{Bottom:} Velocity spectrum $S_\mathrm{EIT}(v_x) = -\alpha_P R^\mathrm{LO}_P(v_x) P_\mathrm{EIT}^\mathrm{th}$. The decrease of its thermal average $S_\mathrm{EIT}^\mathrm{th}$ away from $\kappa_\mathrm{Dopp}$ is seen to be due to a combination of cancelation between $v_x$ populations and spectral peak narrowing.}

\label{fig:doppler1D}
\end{figure}

We study in this paper the adiabatic linear sensitivity $S_\mathrm{EIT}$ defined by
\begin{eqnarray}
P_\mathrm{EIT}[\Omega_\mathrm{Ry}(t)]
&=& P_\mathrm{EIT}(\Omega_\mathrm{LO})
+ S_\mathrm{EIT}(\Omega_\mathrm{LO})
\Omega_\mathrm{in}(t)
\nonumber \\
&&+\ O(|\Omega_\mathrm{in}|^2,
\Delta \omega_\mathrm{in}/\omega_\mathrm{LO})
\label{2}
\end{eqnarray}
in which we identify $S_\mathrm{EIT} = \partial P_\mathrm{EIT}/\partial \Omega_\mathrm{LO}$. As indicated by the error term, even in the linear regime the sensitivity degrades for increasingly off-resonance $|\Delta \omega_\mathrm{in}|$. The rate at which this occurs sets the antenna bandwidth for the given Rydberg state setting \cite{NIST2019c,NIST2022,BBNTS2022}. This will be quantified in detail elsewhere.

The strong illuminations permit a semiclassical modeling approach in which the E-field is treated as classical, coupling to the atom via the standard Stark term, while the spontaneous decays are treated statistically \cite{FIM2005}. Thus, the full atom--photon field problem is replaced by one involving atom states alone, described by a density matrix $\hat \rho$, projected here onto the (five, in this case) illumination driven states. The equation of motion
\begin{equation}
\partial_t \hat \rho = i [\hat \rho, \hat H] + \hat D[\hat \rho]
\label{3}
\end{equation}
(setting $\hbar = 1$) includes both unitary evolution via the Stark Hamiltonian $\hat H$, and nonunitary relaxation via the Lindblad operator $\hat D[\hat \rho]$ that incorporates the spontaneous decay rates \cite{FIM2005}: $D_{mn} = -\frac{1}{2} \rho_{mn} \sum_p (\gamma_{m \to p} + \gamma_{n \to p})$, for $m \neq n$, $D_{nn} = \sum_p (\rho_{pp} \gamma_{p \to n} - \rho_{nn} \gamma_{n \to p})$.

\begin{figure}

\includegraphics[width=3.2in,viewport = 0 0 960 500,clip]{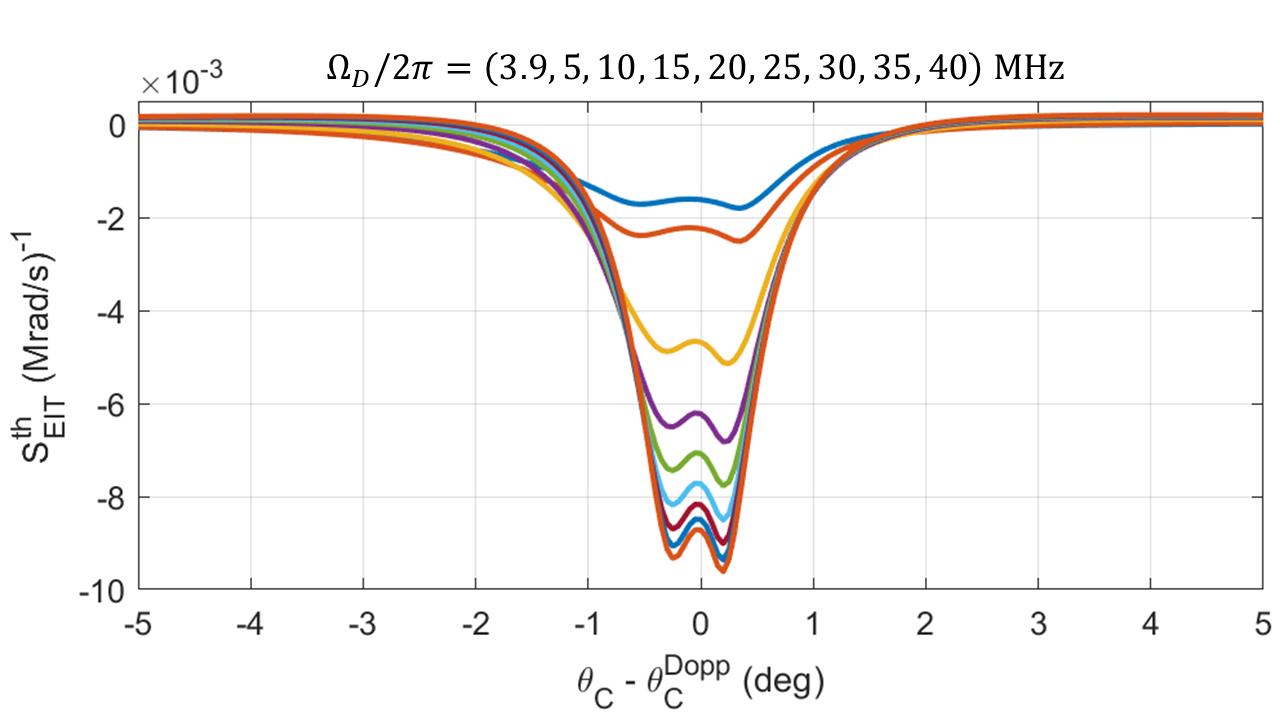}

\caption{EIT sensitivity vs.\ Doppler for the 2D ``star'' setup, pictured in the Fig.\ \ref{fig:SensOmegaLO} inset. The Coupling detuning $\Delta_C + {\bf k}_C \cdot {\bf v}$ is varied through $\theta_C$ relative to $\theta_C^\mathrm{Dopp} = -107.5^\circ$ with fixed $|{\bf k}_C|$. The sensitivity curve maximum magnitudes increase monotonically with Dressing laser amplitude, initially linearly but saturating at $|S_\mathrm{EIT}^\mathrm{th}| \sim 10^{-2}$ for $\Omega_D/2\pi \sim 30$ MHz. Parameters are otherwise the same as in Fig.\ \ref{fig:doppler1D}.}

\label{fig:doppler2D}
\end{figure}

Due to the tree-like structure seen in Fig.\ \ref{fig:rysetup}, in which there are no closed excitation loops (e.g., no additional laser directly coupling states $|3\rangle$, $|5\rangle$ or $|2\rangle$, $|4\rangle$) one may consistently transform to a ``rotating'' frame in which all driving frequencies are absorbed into the states: $|n \rangle \to e^{i(\varepsilon_n + \Delta_n) t} |n\rangle$ and $\rho_{mn} \to e^{i(\varepsilon_m - \varepsilon_n + \Delta_m - \Delta_n) t}\rho_{mn}$, where $\varepsilon_n$ are the bare atom energy levels and $\Delta_n = \varepsilon_n - \varepsilon_m - \omega_{mn}$ is the detuning of the illumination coupling states $|m\rangle$, $|n\rangle$. The Lindblad operator is unaffected and in the absence of the incident field ($\Omega_\mathrm{in} = 0$) the transformed Hamiltonian is time-independent with off-diagonal elements $H_{mn} = -\Omega_{mn}/2$ and diagonal elements given by tree-branch partial sums of detunings. For the five-level system focused on here one obtains
\begin{equation}
H = -\left(\begin{array}{ccc|cc}
0 & \Omega_P/2 & 0 & 0 & 0 \\
\Omega_P^*/2 & \Delta_2 & \Omega_D/2 & 0 & 0 \\
0 & \Omega_D^*/2 & \Delta_3 & \Omega_C/2 & 0 \\ \hline
0 & 0 & \Omega_C^*/2 & \Delta_4 & \Omega_\mathrm{LO}/2 \\
0 & 0 & 0 & \Omega_\mathrm{LO}^*/2 & \Delta_5
\end{array} \right)
\label{4}
\end{equation}
in which $\Delta_2 = \Delta_P$, $\Delta_3 = \Delta_P + \Delta_D$, $\Delta_4 = \Delta_P + \Delta_D + \Delta_C$, $\Delta_5 = \Delta_P + \Delta_D + \Delta_C + \Delta_\mathrm{LO}$. The lines emphasize the core and Rydberg subspaces. For atom velocity ${\bf v}$, the energy levels are Doppler shifted relative to the stationary illuminators, with resulting detuning shifts $\Delta_\alpha \to \Delta_\alpha + {\bf k}_\alpha \cdot {\bf v}$ where ${\bf k}_\alpha$ is the wavevector of illuminator $\alpha$. Note that even as $v/\lambda_\mathrm{LO}$ is at least $O(10^4)$ smaller than the laser values, ${\bf E}_\mathrm{LO}({\bf x})$ will also not be plane-wave like, being strongly influenced by the vapor cell geometry. The Doppler shift of $\Delta_\mathrm{LO}$ will hence be neglected.

\begin{figure*}

\includegraphics[width=2.3in,viewport = 0 0 660 560,clip]{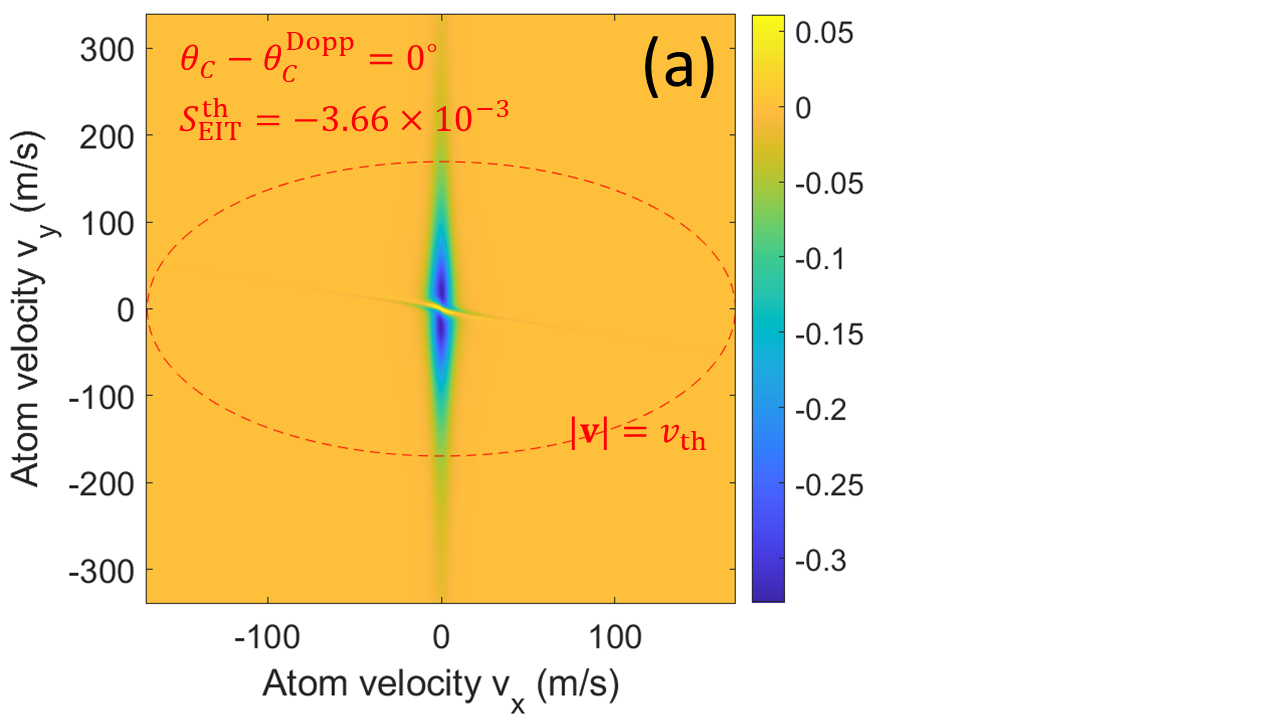}
\includegraphics[width=2.3in,viewport = 0 0 660 560,clip]{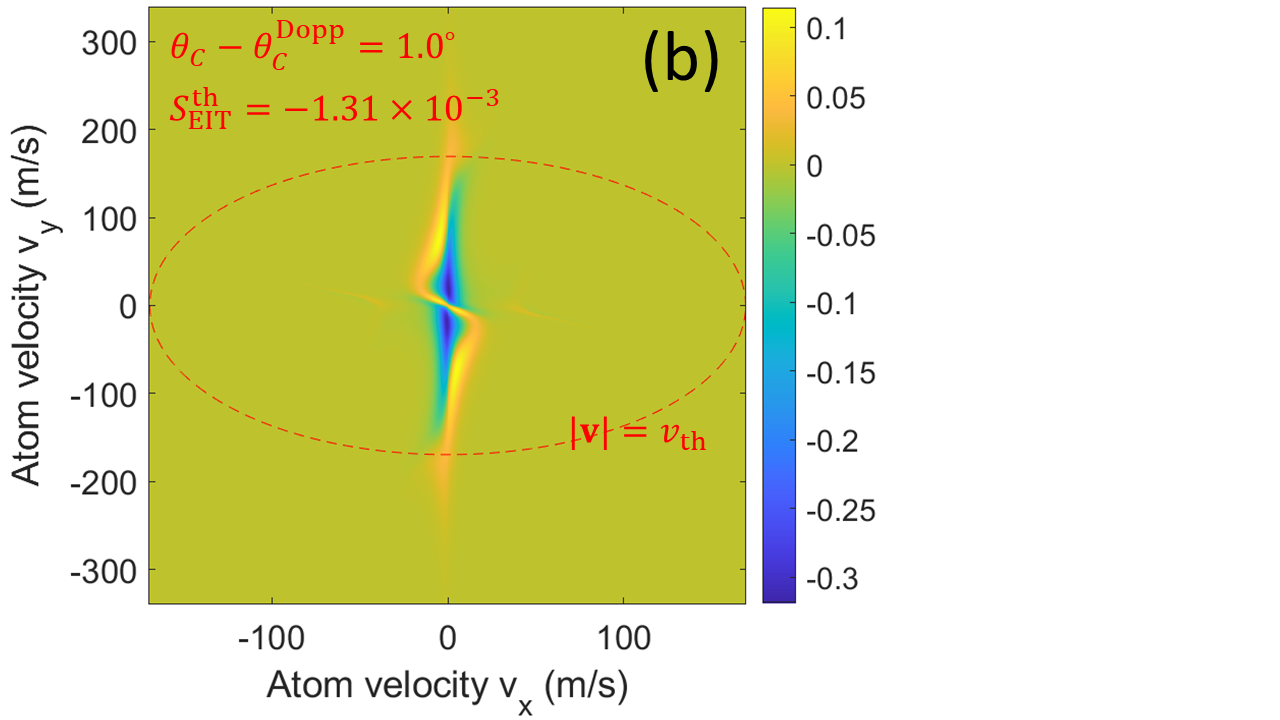}
\includegraphics[width=2.3in,viewport = 0 0 660 560,clip]{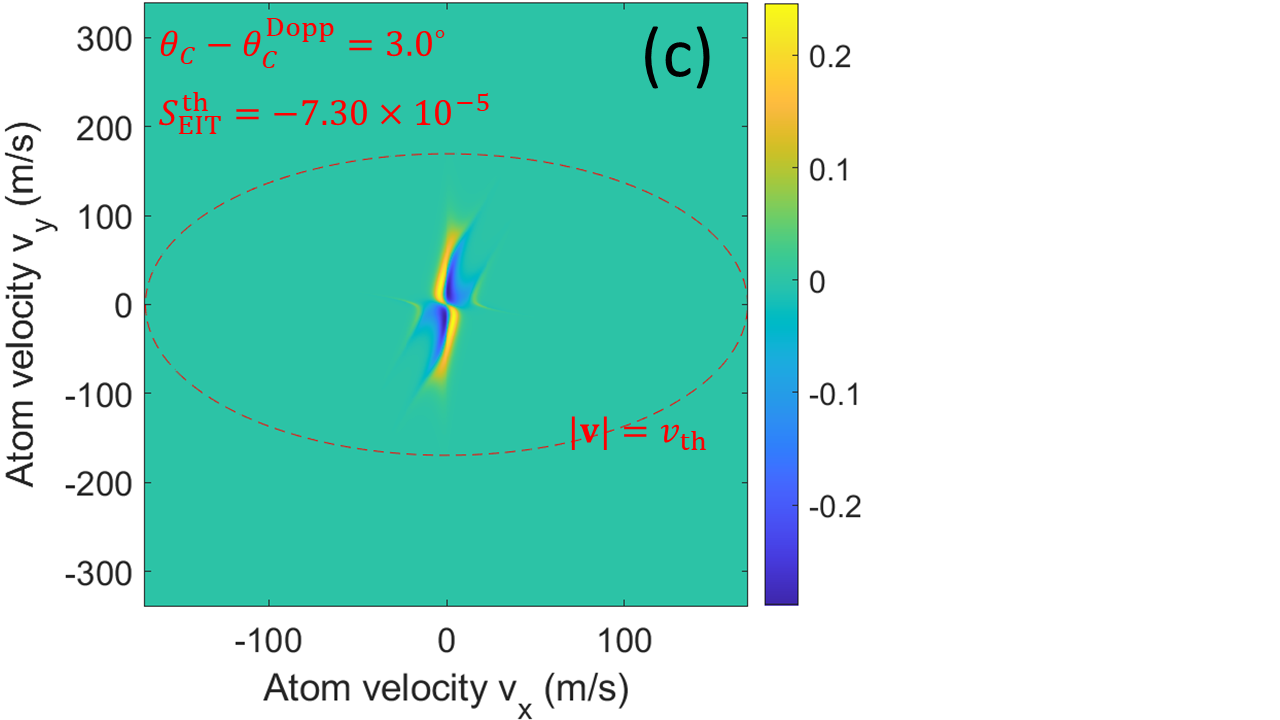}

\caption{EIT sensitivity velocity spectra $S_\mathrm{EIT}({\bf v}) = -\alpha_P R^\mathrm{LO}_P({\bf v}) P_\mathrm{EIT}^\mathrm{th}$ corresponding to the $\Omega_\mathrm{LO}/2\pi = 0.85$ MHz case in Fig.\ \ref{fig:doppler2D}. The magnitude of the thermal averages (\ref{7}), $S^\mathrm{th}_\mathrm{EIT} = [-36.6 \mbox{ \textbf{(a)}}, -13.1 \mbox{ \textbf{(b)}}, -0.730 \mbox{ \textbf{(c)}}] \times 10^{-4}$ (Mrad/s)$^{-1}$, decrease rapidly away from the Doppler angle $\theta_C^\mathrm{Dopp} = -107.5^\circ$ (Fig.\ \ref{fig:SensOmegaLO} inset) even while the color scale remains basically unchanged.}

\label{fig:vspec2d}
\end{figure*}

Since the (\ref{3}) is linear in $\hat \rho$, by listing its elements as a column vector ${\bm \rho}$, one obtains
\begin{equation}
\partial_t {\bm \rho} = {\bf G} {\bm \rho},\ \
{\bf G} = i{\bf H} + {\bf D}
\label{5}
\end{equation}
with commutator matrix $H_{mn,pq} = H_{qn} \delta_{pm} - H_{mp} \delta_{qn}$ and elements of ${\bf D}$ similarly determined by the decay rates. The steady state, or adiabatic, density matrix ${\bm \rho}_\mathrm{ad}$ therefore satisfies ${\bf G}{\bm \rho}_\mathrm{ad} = 0$. Existence of a nonempty kernel follows from conservation of probability, ${\bf P} \cdot {\bm \rho} \equiv \mathrm{tr}[\hat \rho] = 1$, translating to ${\bf P}^T {\bf G} = 0$, guaranteeing reduced rank of ${\bf G}$. The sensitivity derivative, as seen below, will involve the same derivative $\partial {\bm \rho}/\partial\Omega_\mathrm{LO}$. One obtains
\begin{equation}
\partial_a {\bm \rho}_\mathrm{ad} = -{\bf G}^{-1}
(\partial_a {\bf G}) {\bm \rho}_\mathrm{ad},\
{\bf G}^{-1} \equiv \sum_{\lambda_n \neq 0}
\frac{1}{\lambda_n} {\bf u}^L_n {\bf u}^{R \dagger}_n
\label{6}
\end{equation}
where $a$ is any parameter, and ${\bf G}^{-1}$ is the restricted inverse computed as shown from its nonzero eigenvalues $\lambda_n$ and left and right eigenvectors ${\bf u}^{L,R}_n$, with normalization ${\bf u}^{L \dagger}_m {\bf u}_n^R = \delta_{mn}$. For any illumination parameter, one obtains the rather sparse matrix $\partial_a {\bf G} = i\partial_a {\bf H}$.

Thermal averages are performed using the Maxwell distribution (appropriate to finite $T$ dilute vapors):
\begin{equation}
\hat \rho^\mathrm{th} = \left(\frac{m}{2\pi k_B T} \right)^{3/2}
\int d{\bf v} e^{-m{\bf v}^2/2k_BT} \hat \rho({\bf v})
\label{7}
\end{equation}
with atomic mass  $m$. For $T = 300$ K the $^{87}$Rb thermal velocity is $v_\mathrm{th} = \sqrt{k_B T/m} = 169$ m/s. For $\lambda = 1\ \mu$m this leads to Doppler shifts $v_\mathrm{th}/\lambda \simeq 170$ MHz, enormous compared to the $\sim$1 MHz, or less, resonant linewidths encountered below. A naive estimate is hence that fewer than 1\% and 0.01\% of atoms, that happen to be moving slowly, will contribute to $S^\mathrm{th}_\mathrm{EIT}$ for 1D and 2D setups, respectively. In fact, it is generally much worse than this. It will be seen that the ${\bf v}$ dependence causes different atom populations to give opposite-sign, near-canceling contributions to $S^\mathrm{th}_\mathrm{EIT}$ unless one enforces the condition $\sum_\alpha {\bf k}_\alpha = 0$. The latter \emph{does not} lead to true Doppler compensation (e.g., effectively smaller $T$), it simply restores the above naive Doppler-reduced estimate.

The EIT response is derived from $\hat \rho^\mathrm{th}$ in the form \cite{FIM2005}
\begin{eqnarray}
P^\mathrm{th}_\mathrm{EIT}(L) &=& \frac{\Omega_P(L)^2}{\Omega_P(0)^2}
= e^{-\alpha_P R_P(L)}
\nonumber \\
S^\mathrm{th}_\mathrm{EIT}(L) &=& -\alpha_P
R^\mathrm{LO}_P(L) P^\mathrm{th}_\mathrm{EIT}(L),\ \
R_P^\mathrm{LO} \equiv \frac{\partial R_P(L)}{\partial \Omega_\mathrm{LO}}
\nonumber \\
R_P(L) &\equiv& \int_0^L \frac{\mathrm{Im}[\rho^\mathrm{th}_{21}(s)]}{\Omega_P(s)} ds
\label{8}
\end{eqnarray}
with cell length $L$, $\alpha_P = 2k_P N_0 |{\bf d}_{12}|^2/\epsilon_0 \hbar$, and atomic number density $N_0$. This is a linear response result, failing for large $|\Omega_P|$ (hence competing with the desire to increase photon count); $\Omega_P/2\pi < 3$ MHz is found to be a reasonable experimental compromise. Note that $\mathrm{Re}[\rho^\mathrm{th}_{21}]$ corresponds to an index of refraction, perhaps detectable via an alternative interference measurement. Note also that absorption is a consequence of decay processes: $\mathrm{Im}[\hat \rho^\mathrm{th}]$ is nonzero only due to the presence of $\hat D[\hat \rho^\mathrm{th}]$, mainly through the relatively large value $\gamma_{12}$. Spectral properties of $\hat H$ dominate the resonant behavior, but $\hat D[\hat \rho^\mathrm{th}]$ generates the EIT signal.

If one assumes that only $\Omega_P$ varies along the beam, then (\ref{8}) is computed by solving the ODE pair
\begin{eqnarray}
\partial_s \Omega_P(s) &=& -\frac{1}{2} \alpha_P
\mathrm{Im}[\rho^\mathrm{th}_{21}[\Omega_P(s)]]
\nonumber \\
\partial_s R_P^\mathrm{LO}(s)
&=& \frac{1}{\Omega_P(s)}
\mathrm{Im}\left[\frac{\partial\rho^\mathrm{th}_{21}}{\partial \Omega_{LO}}
[\Omega_P(s)] \right]
\label{9} \\
&-& \frac{\alpha_P}{2}
\Omega_P(s) R_P^\mathrm{LO}(s)
\mathrm{Im}\left[\frac{\partial}{\partial \Omega_P(s)}
\frac{\rho^\mathrm{th}_{21}[\Omega_P(s)]}{\Omega_P(s)} \right]
\nonumber
\end{eqnarray}
with all other parameters taken as fixed. The two density matrix derivatives are computed via (\ref{6}). Depending on setup details, these could be extended to treat simultaneous inhomogeneity of the remaining $\Omega_\alpha$, generated by exponentials of other components of $\hat \rho^\mathrm{th}$.

\begin{figure}

\includegraphics[width=3.0in,viewport = 0 0 780 540,clip]{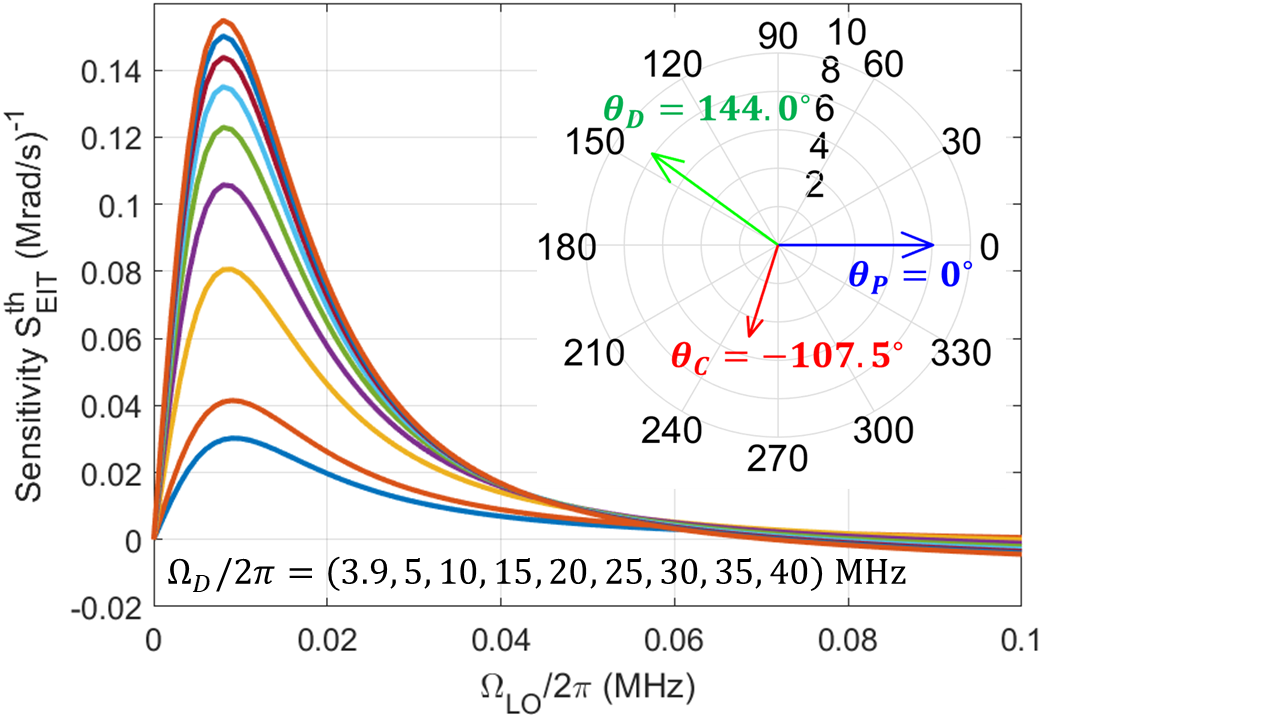}

\caption{EIT sensitivity in the regime of very small local oscillator strength $\Omega_\mathrm{LO}$. The $O(10\ \mathrm{kHz})$ linewidth implies a similar limitation on the linear response regime. The negative $S_\mathrm{EIT}^\mathrm{th}$ values at $\Omega_\mathrm{LO}/2\pi = 0.85$ MHz seen in Fig.\ \ref{fig:doppler2D} reach a local minimum at $\Omega_\mathrm{LO}/2\pi \sim 0.4$ MHz, reverse sign at $\Omega_\mathrm{LO}/2\pi \sim 0.1$ MHz as seen here, then grow to much larger positive values. The enhancement with increasing $\Omega_D$ seen in Fig.\ \ref{fig:doppler2D} is preserved. \textbf{Inset:} Star configuration ($k_\alpha$ units of rad/nm).}

\label{fig:SensOmegaLO}
\end{figure}

For $L \alt 1$ cm we find that $\Omega_P \simeq \Omega_P(0)$ may be taken as uniform, yielding the simplification $R_P(L) = (L/\Omega_P) \mathrm{Im}[\rho^\mathrm{th}_{21}(\Omega_P)]$, and leading to
\begin{equation}
S^\mathrm{th}_\mathrm{EIT}(L) = -\frac{\alpha_P L}{\Omega_P}
 \mathrm{Im}\left[\frac{\partial \rho^\mathrm{th}_{21}(\Omega_P)}
{\partial\Omega_\mathrm{LO} }\right]
e^{-\alpha_P R_P(L)}.
\label{10}
\end{equation}
For numerical simplicity, most examples below will use this limit. At the end we consider solutions of (\ref{9}) for larger $L$, addressing questions of optimal cell length.

The first (1D setup) Doppler effect demonstration is shown in Fig.\ \ref{fig:doppler1D}. The maximum sensitivity $S^\mathrm{th}_\mathrm{EIT} \sim 10^{-3}$ (Mrad/s)$^{-1}$, indeed about 1\% of the $T=0$ value (not shown), occurs very close to $(k_C/k_P)^\mathrm{Dopp} \equiv 1 - k_D/k_P = -0.00549$, and drops by $O(10^2)$ at the physical value $k_C/k_P = 0.6197$ in Fig.\ \ref{fig:rysetup}. Since $k_C$ cannot physically be varied without strongly varying $\Delta_C$, this result is artificial, but is next confirmed for the physically consistent 2D ``star'' configuration ${\bf k}_P + {\bf k}_D + {\bf k}_C = 0$ (Fig.\ \ref{fig:SensOmegaLO} inset). Figure \ref{fig:doppler2D} shows the effect of perturbation of the angle $\theta_C$ of ${\bf k}_C$ for various $\Omega_D$. The sensitivity magnitude in all cases drops by $O(10)$ on $\sim 1^\circ$ scales, corresponding to $\sim$1 MHz shift in the detuning $\Delta_C$. The $O(10^{-4})$ peak sensitivity value for $\Omega_D/2\pi = 3.9$ MHz is consistent with the naive estimate above. The increase with $\Omega_D$ (and with $\Omega_C$, but not shown) occurs because increasing the off-diagonal components of the Hamiltonian (\ref{4}) reduces the influence of the Doppler-dominated diagonal components. The effect saturates above $\Omega_D/2\pi \sim 30$ MHz because the increasing velocities involved are suppressed by the Maxwell distribution (\ref{7}).

Insight into these effects is obtained from the underlying 1D and 2D ``velocity spectra'' shown, respectively, in the bottom panel of Fig.\  \ref{fig:doppler1D} and in Fig.\ \ref{fig:vspec2d}. Different atom populations clearly contribute very differently. For example, for the 2D case with $\theta_C = \theta_C^\mathrm{Dopp}$, the stationary atom value $S_\mathrm{EIT}({\bf v} = 0) \simeq 0.05$ is actually dominated by neighboring large negative sensitivity regions $S_\mathrm{EIT}({\bf v}) \sim -0.3$. In both 1D and 2D cases, away from the Doppler point, the scale of variation with ${\bf v}$ is unchanged, but the geometry of the positive and negative regions changes dramatically, with $S_\mathrm{EIT}^\mathrm{th} \to 0$ through a combination of cancellation between regions and shrinking of the near-resonant region.

Note that even as the 1D configuration has a built in $O(10^{-2})$ Doppler disadvantage, it is simpler to set up and has geometric advantages, such as larger $L$ and conveniently overlapping laser beams, that have made it the dominant experimental focus. The active vapor volume could then be increased to gain a factor of $O(10)$ over the result in Fig.\ \ref{fig:doppler1D}. Similar 2D setup gains the would require greatly broadening the $\sim$1 mm$^2$ intersecting beams.

\begin{figure}

\includegraphics[width=3.0in,viewport = 0 0 960 500,clip]{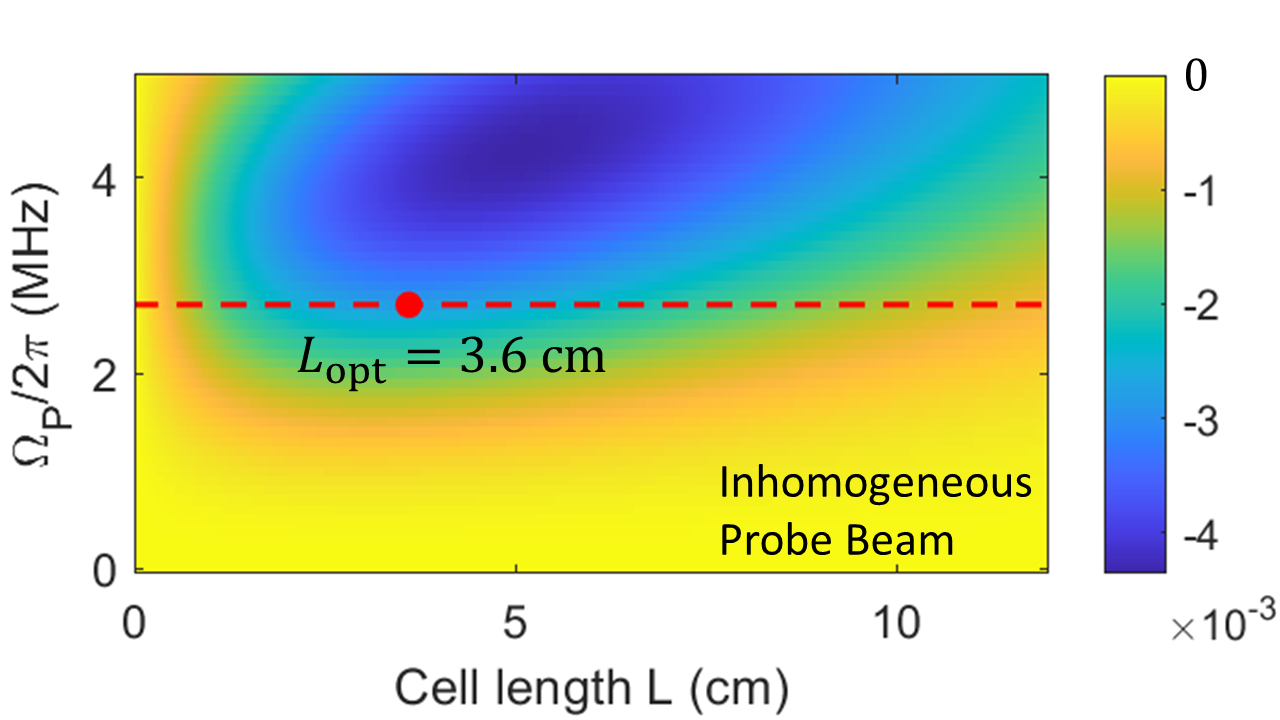}
\includegraphics[width=3.0in,viewport = 0 0 960 500,clip]{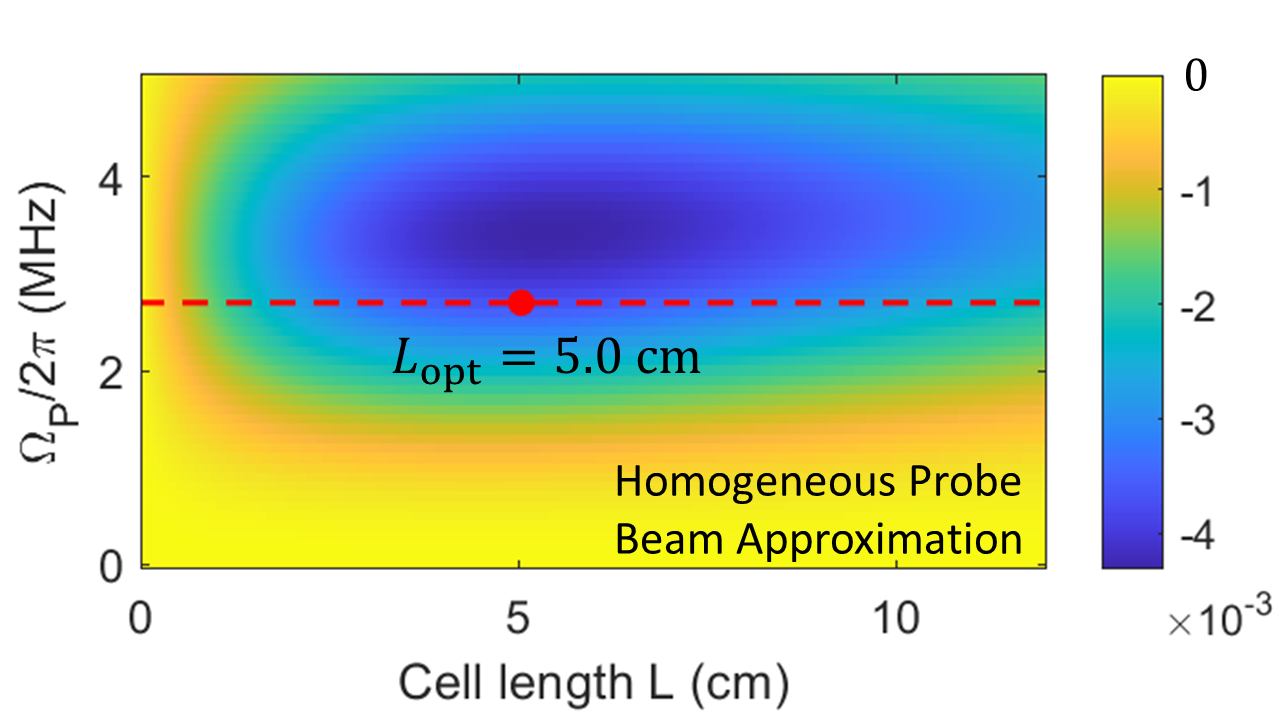}

\caption{Sensitivity (Mrad/s)$^{-1}$ vs.\ Probe beam amplitude $\Omega_P \equiv \Omega_P(0)$ and cell length, limiting $\Omega_P/2\pi \leq 5$ MHz so as not to stray too far out of the estimated linear response regime. Parameters are otherwise as in Fig.\ \ref{fig:doppler1D}. \textbf{Above:} Full inhomogeneous solution (\ref{8}) using $\Omega_P(s)$ obtained from (\ref{9}). \textbf{Below:} Homogeneous approximation using uniform $\Omega_P(s) = \Omega_P$. The red dashed line highlights the result for $\Omega_P/2\pi = 2.7$ MHz for which rather different optimal lengths (red dots) are found.}

\label{fig:SenseInhomog}
\end{figure}

Figure \ref{fig:SensOmegaLO} shows an exploration of the very small LO amplitude limit (now limited to $\theta_C = \theta_C^\mathrm{Dopp}$). For vanishing $\Delta_\alpha$ used here, symmetry dictates that $P^\mathrm{th}_\mathrm{EIT}$ be even, hence $S^\mathrm{th}_\mathrm{EIT} \to 0$ for $\Omega_\mathrm{LO} \to 0$. Indeed, a negative maximum is found for $\Omega_\mathrm{LO}/2\pi \sim 0.4$ MHz, but pushing to lower $\Omega_\mathrm{LO}$ yields the surprising result that  $S_\mathrm{EIT}$  changes sign at $\Omega_\mathrm{LO}/2\pi \sim 0.1$ MHz, then grows to an $O(10^2)$ times larger positive maximum at $\Omega_\mathrm{LO}/2\pi \sim 10$ kHz, before finally turning around to vanish. This maximum also grows roughly linearly with $\Omega_D$.

Finally we return to consider inhomogeneous probe beams. Figure \ref{fig:SenseInhomog} shows exact and approximate sensitivity results vs.\ input power and cell length. The differences are not huge, but for given $\Omega_P(0)$ the two can give very different predictions for the maximum sensitivity point, highlighted here for $\Omega_P(0)/2\pi = 2.7$ MHz. In either case, one can more than double the sensitivity by going to considerably longer than $L = 1$ cm. Of course there are practical issues that may make this difficult.

We have explored here several avenues for increasing Rydberg antenna sensitivity using 2D ``Doppler aware'' setups. The highlight, perhaps, is demonstration of an $O(10^2)$ boost from $S_\mathrm{EIT}^\mathrm{th} \sim 10^{-3}$ (upper curve of Fig.\ \ref{fig:doppler2D}, representing ``typical'' experimental setups) to $\sim 10^{-1}$ (upper curve of  Fig.\ \ref{fig:SensOmegaLO}, optimally small $\Omega_\mathrm{LO}$, large $\Omega_D$). There are a number of extensions of the theory, relevant to applications, that will be presented elsewhere: measurement noise considerations (e.g., photon shot noise replaces $S_\mathrm{EIT}$ by the signal-to-noise ratio measure $\mathrm{SNR} = \Omega_P S^\mathrm{th}_\mathrm{EIT}/\sqrt{P^\mathrm{th}_\mathrm{EIT}}$); non-adiabatic effects and accompanying signal bandwidth limitations; as well as other forms of signal distortion including nonlinear response. The five-state projection (\ref{4}) also deserves more careful consideration. Influence of additional nearby but nonresonant states is unlikely to change the basic conclusions, but could influence detailed line shapes in the high resolution limits of interest here.

\emph{Acknowledgments:} This material is based upon work supported by the Defense Advanced Research Projects Agency (DARPA) under Contract No.\ HR001121C0122. The author also benefited from numerous conversations with Craig Price, Ying Ju Wang, Eric Bottomley, Haoquan Fan, and Shane Verploegh regarding experimental considerations.

\end{document}